\documentclass[]{aastex}
\slugcomment{Accepted for publication in the Astrophysical Journal Letters}
\begin{document}
\title{Dense optical and near-infrared monitoring of CTA 102 during high state in 2012 with {\it OISTER}: Detection of intra-night ``orphan polarized flux flare"}
\author{
Ryosuke Itoh\altaffilmark{1,2}, 
Yasushi Fukazawa\altaffilmark{1}, 
Yasuyuki T. Tanaka\altaffilmark{1}, 
Yuhei Abe\altaffilmark{3}, 
Hiroshi Akitaya\altaffilmark{4}, 
Akira Arai\altaffilmark{5}, 
Masahiko Hayashi\altaffilmark{6}, 
Takafumi Hori\altaffilmark{7}, 
Mizuki Isogai\altaffilmark{8}, 
Hideyuki Izumiura\altaffilmark{9}, 
Koji S. Kawabata\altaffilmark{4}, 
Nobuyuki Kawai\altaffilmark{10}, 
Daisuke Kuroda\altaffilmark{9}, 
Ryo Miyanoshita\altaffilmark{11}, 
Yuki Moritani\altaffilmark{4}, 
Tomoki Morokuma\altaffilmark{12}, 
Takahiro Nagayama\altaffilmark{13}, 
Jumpei Nakamoto\altaffilmark{14}, 
Chikako Nakata\altaffilmark{7}, 
Yumiko Oasa\altaffilmark{15}, 
Tomohito Ohshima\altaffilmark{7}, 
Takashi Ohsugi\altaffilmark{4}, 
Shin-ichiro Okumura\altaffilmark{16}, 
Yoshihiko Saito\altaffilmark{10}, 
Yu Saito\altaffilmark{17}, 
Mahito Sasada\altaffilmark{7}, 
Kazuhiro Sekiguchi\altaffilmark{6}, 
Yuhei Takagi\altaffilmark{5}, 
Jun Takahashi\altaffilmark{5}, 
Yukihiro Takahashi\altaffilmark{3,14}, 
Katsutoshi Takaki\altaffilmark{1}, 
Makoto Uemura\altaffilmark{4}, 
Issei Ueno\altaffilmark{1}, 
Seitaro Urakawa\altaffilmark{16}, 
Makoto Watanabe\altaffilmark{3,14}, 
Masayuki Yamanaka\altaffilmark{18}, 
Yoshinori Yonekura\altaffilmark{19} and 
Michitoshi Yoshida\altaffilmark{4}
}
\altaffiltext{1}{Department of Physical Sciences, Hiroshima University, Higashi-Hiroshima, Hiroshima 739-8526, Japan}
\altaffiltext{2}{itoh@hep01.hepl.hiroshima-u.ac.jp}
\altaffiltext{3}{Department of Cosmosciences, Graduate School of Science, Hokkaido University, Kita-ku, Sapporo 060-0810, Japan}
\altaffiltext{4}{Hiroshima Astrophysical Science Center, Hiroshima University, Higashi-Hiroshima, Hiroshima 739-8526, Japan}
\altaffiltext{5}{Nishi-Harima Astronomical Observatory, Center for Astronomy, University of Hyogo, 407-2, Nishigaichi, Sayo-cho, Sayo, Hyogo 679-5313, Japan}
\altaffiltext{6}{National Astronomical Observatory of Japan, Osawa 2-21-2, Mitaka, Tokyo, 181-8588, Japan}
\altaffiltext{7}{Department of Astronomy, Graduate School of Science, Kyoto University, Sakyo-ku, Kyoto 606-8502, Japan}
\altaffiltext{8}{Koyama Astronomical Observatory, Kyoto Sangyo University, Motoyama, Kamigamo, Kita-Ku, Kyoto-City 603-8555, Japan}
\altaffiltext{9}{Okayama Astrophysical Observatory, National Astronomical Observatory of Japan, Honjo 3037-5, Kamogata, Asakuchi, Okayama 719-0232, Japan}
\altaffiltext{10}{Department of Physics, Tokyo Institute of Technology, 2-12-1 Ookayama, Meguro-ku, Tokyo 152-8551, Japan}
\altaffiltext{11}{Graduate School of Science and Engineering, Kagoshima University, 1-21-35 Korimoto, Kagoshima 890-0065, Japan}
\altaffiltext{12}{Institute of Astronomy, Graduate School of Science, The University of Tokyo, 2-21-1 Osawa, Mitaka, Tokyo 181-0015, Japan}
\altaffiltext{13}{Department of Astrophysics, Nagoya University, Chikusa-ku Nagoya 464-8602, Japan}
\altaffiltext{14}{Department of Earth and Planetary Sciences, School of Science, Hokkaido University, Kita-ku, Sapporo 060-0810, Japan}
\altaffiltext{15}{Faculty of Education, Saitama University, 255 Shimo-Okubo, Sakura, Saitama, 338-8570, Japan}
\altaffiltext{16}{Bisei Spaceguard Center, Japan Spaceguard Association, 1716-3 Okura, Bisei-cho, Ibara-shi, Okayama 714-1411, Japan}
\altaffiltext{17}{College of Science, Ibaraki University, 2-1-1 Bunkyo, Mito, Ibaraki 310-8512, Japan}
\altaffiltext{18}{Kwasan Observatory, Kyoto University, Ohmine-cho Kita Kazan, Yamashina-ku, Kyoto 607-8471, Japan}
\altaffiltext{19}{Center for Astronomy, Ibaraki University, 2-1-1 Bunkyo, Mito, Ibaraki 310-8512, Japan}

\begin{abstract}
CTA~102, classified as a flat spectrum radio quasar at z=1.037, produced exceptionally bright optical flare in 2012 September. Following {\it Fermi}-LAT detection of enhanced
$\gamma$-ray activity, we densely monitored this source in the optical and near-infrared bands for the subsequent ten nights using twelve telescopes in Japan and South-Africa. On MJD 56197 (2012 September 27, 4-5 days after the peak of bright $\gamma$-ray flare), polarized flux showed a transient increase, while total flux and polarization angle remained almost constant during the ``orphan polarized-flux flare". We also detected an intra-night and prominent flare on MJD 56202. The total and polarized fluxes showed quite similar temporal variations, but PA again remained constant during the flare. Interestingly, the polarization angles during the two flares were significantly different from the jet direction. Emergence of a new emission component with high polarization degree (PD) up to 40\% would be responsible for the observed two flares, and such a high PD indicates a presence of highly ordered magnetic field at the emission site. We discuss that the well-ordered magnetic field and even the observed directions of polarization angle which is grossly perpendicular to the jet are reasonably accounted for by transverse shock(s) propagating down the jet.
\end{abstract}
\maketitle
\section{Introduction}
Blazars are highly variable active galactic nuclei
(AGN) emitting radiation at all wavelengths from
radio to $\gamma$-rays. They have strong
relativistic jets aligned with the observer's line
of sight and are apparently bright due to
relativistic beaming. Their emission typically
consists of two spectral components. One is
attributed to synchrotron radiation at lower
energies peaking in the radio through optical bands,
and the other is inverse Compton scattering peaking
in the $\gamma$-ray bands. Outstanding
characteristics of blazars are their rapid and
high-amplitude intensity variations or flares. These
variabilities are observed in various wavelengths
and time scales. Micro variability (intra-day
variability) of flux in the optical band has been
also detected on timescales as short as minutes to
hours \citep[e.g., ][]{1970ApJ...159L..99R,
  1996ASPC..110...17M}. It is important to measure
the time scales of micro variability because it
provides limits on the size and location of the
emitting regions.

Polarized radiation is one of the evidences of synchrotron origin in low energies and it also varies drastically.
Therefore, optical polarimetric observations also provide a strong tool to probe
jet structures \citep[e.g., ][]{2008Natur.452..966M, 2010Natur.463..919A}.
Nevertheless, simultaneous short-term (intra-day) observations of flux, color and polarization 
have been performed only in a few blazars 
\citep[e.g., ][]{2003A&A...409..857A, 2007MNRAS.381L..60C, 2008PASJ...60L..37S, 2008ApJ...672...40H, 2011A&A...531A..38A},
and hence the origin of micro variability is still unclear.

CTA~102 (also known as PKS J2232+1143, R.A. = $22^h 32^m 36.4^s$,
decl. = $+11\arcdeg\ 43\arcmin\ 50\arcsec.8$, J2000, z=1.037, \citealt{1989ApJS...69....1H})
was first identified as a strong radio source \citep{1960PASP...72..237H} 
and classified as a flat spectrum radio quasar \citep[FSRQ,][]{2010ApJ...722..520A} from multi-wavelength observations.
CTA~102 showed micro variability of optical flux and color in the 2004 flare \citep{2009AJ....138.1902O}.
In this flare, the ``redder when brighter'' trend was reported and 
it was able to be explained by the superposition of 
variable synchrotron component from radio to optical bands and non-variable ``blue bump'' 
component which is thought to be thermal disk radiation connecting to the UV band.

Recently, CTA~102 showed an extreme activity in the
optical and GeV $\gamma$-ray bands on September 2012
\citep{2012ATel.4397....1L,2012ATel.4409....1O}.  
In this paper, we present results of
high-temporal-density monitoring observations of
CTA~102 just after the September 2012 $\gamma$-ray
flare.

\section{Observation}

\begin{table*}
  \centering
  \caption{Observatories, Telescopes and Instruments}
  \label{tab:teles}
 {\small
  \begin{tabular}{lrcc}\hline\hline
    Observatory/Telescope   & diameter\tablenotemark{a} & Instrument    &  Filters   \\ \hline
    \multicolumn{4}{c}{Opt/NIR in Japan} \\ \hline
    Nayoro Observatory/Pirka                       & 160 cm & MSI$^{(1)}$      & {\it V, R$_{\rm{C}}$} \\
    Akeno Observatory/MITSuME$^{(2)}$               &  50 cm &                 & {\it g', R$_{\rm{C}}$, I$_{\rm{C}}$}          
     \\
    Kyoto University/--                            &  40 cm &                 & {\it R$_{\rm{C}}$}                     \\
    Koyama Astronomical Observatory/Araki          & 130 cm & ADLER           & 
{\it B, g', V, I$_{\rm{C}}$, i', z'}  \\
    Nishi-Harima Astronomical Observatory/Nayuta   & 200 cm & NIC             & {\it K$_{\rm{s}}$}                 \\    
    Bisei Spaceguard Center/--                     & 100 cm & Volante         & {\it r'}                    \\
    Okayama Astrophysical Observatory/--           & 188 cm & ISLE$^{(3,4)}$            & {\it J, H, K$_{\rm{s}}$}           \\
    Okayama Astrophysical Observatory/MITSuME$^{(2)}$ &  50 cm &                 & {\it g', R$_{\rm{C}}$, I$_{\rm{C}}$}               \\
    Higashi-Hiroshima Observatory/Kanata           & 150 cm & HOWPol$^{(5)}$   & {\it V, R$_{\rm{C}}$, R$_{\rm{C}}$}-Pol.   \\
    Iriki Observatory/--                           & 100 cm & Infrared Camera & {\it J, H, K'}              \\ 
    \hline
    \multicolumn{4}{c}{Opt/NIR in South-Africa} \\ \hline
    South African Astronomical Observatory/IRSF     & 140 cm & SIRIUS          & {\it J, H, K$_{\rm{s}}$}           \\ 
    \hline
    \multicolumn{4}{c}{Radio Observatory} \\ \hline
     Mizusawa VLBI Observatory/Hitachi 32-m Telescope                   & 32 \ m  &             & 8.4 GHz                 \\ \hline
  \end{tabular}
  }
  \tablerefs{(1) \citealt{2012SPIE.8446E..2OW};
    (2) \citealt{2005NCimC..28..755K};
    (3) \citealt{2006SPIE.6269E.118Y};
    (4) \citealt{2008SPIE.7014E.106Y};
    (5) \citealt{2008SPIE.7014E.151K}.}
  \begin{flushleft} 
  ${}^a$ Size of primary mirror
  \end{flushleft}
\end{table*}

Observations were carried out as a ToO program of Optical and Infrared Synergetic Telescopes for Education and Research ({\it OISTER}). 
{\it OISTER} is a global observing network that
links organically many ground-based small telescopes
in Japan, South Africa, and Chile under a Japanese interuniversity cooperation regime. 
{\it OISTER} is aimed at investigating potential transient sources 
($\gamma$-ray bursts, AGNs, supernovae, cataclysmic variables, and so on). 
The biggest advantage of {\it OISTER} is its capability of performing continuous and
high-temporal-density monitoring in many bands extending to as long a wavelength as {\it K$_{\rm{s}}$}.
ToO observation of CTA~102 with {\it OISTER} was conducted from
September 23 to October 3 in 2012, following the bright GeV $\gamma$-ray flare.
We obtained the {\it B, V, g', r', R$_{\rm{C}}$, I$_{\rm{C}}$, i', z', J, H, K'} and {\it K$_{\rm{s}}$} band photometric and
the {\it R$_{\rm{C}}$} band polarimetry data with {\it OISTER} and also with other collaborative telescopes.
Telescopes and instruments used for this observation are listed on Table \ref{tab:teles}.
Note that we treated the {\it r'} and {\it K'} band data as the {\it R$_{\rm{C}}$} and {\it K$_{\rm{s}}$} band data, respectively.
Reductions of optical and near-infrared (NIR) data were performed under the
standard procedure of CCD photometry.
The position of comparison star is
R.A. = $22^h 32^m 41.5^s$, decl. = $+11\arcdeg\ 43\arcmin\ 14\arcsec.1$ (J2000).
The magnitudes in the optical bands were obtained differentially with nearby comparison 
stars which have been calibrated with the photometric standard stars in Landolt fields 
\citep{1992AJ....104..340L} observed on clear and stable nights.
The same comparison stars are used also in NIR bands,
where the magnitude is given in the {\it 2MASS} catalogue \citep{2006AJ....131.1163S}.
We corrected the data for the Galactic extinction
\citep[e.g., $A_{\rm V}$=0.233,][NED
  database\footnote{http://ned.ipac.caltech.edu/}]{2011ApJ...737..103S}.
There were small systematic difference in photometric system among observatories and instruments; the standard deviations of the
magnitudes of the comparison star during the observation period were 
$\Delta R_{\rm{C}}\sim 0.02$ mag ($\sim 2\%$ of flux) and $\Delta K_{\rm{s}}\sim 0.06$ mag  ($\sim 5\%$ of flux). 
These values were added to the photometric errors of CTA~102 in each band.

The polarimetric observations were performed with HOWPol installed to the Kanata Telescope 
located on Higashi-Hiroshima Observatory \citep{2008SPIE.7014E.151K}.
A unit of the observing sequence consisted of successive exposures
at 4 position angles of a half-wave plate;
$0^{\circ}, 45^{\circ}, 22.5^{\circ}$ and $67.5^{\circ}$.
Polarimetry with HOWPol suffers from large instrumental polarization ($\Delta p\sim 4$\%) 
produced by the reflection of the incident light on the tertiary mirror of the telescope.
The instrumental polarization was modeled as a function of the declination of the object and the 
hour angle at the observation, and we subtracted it from the observation.
We estimated that the error in this instrumental polarization
correction is smaller than 0.5\% from many observations for
unpolarized standard stars.
The polarization angle (PA) is defined in the
standard manner as measured from north to east.
The PA was calibrated with two polarized stars, HD183143 and HD204827
\citep{1983A&A...121..158S}.
Because the PA has an ambiguity of $\pm 180^{\circ} \times
n$ (where $n$ is an integer),
we selected $n$ which gives the least angle
difference from the previous data, assuming that
the PA would change smoothly.
The error of PA was estimated to be smaller than $2^{\circ}$ from observations of the polarized stars.

The radio data were obtained from September 26 to 28
using Hitachi 32 m telescope of Mizusawa VLBI
Observatory, NAOJ, which is operated by Ibaraki
University. The front end was a cooled HEMT
receiver, and a typical system temperature was 25 K
including the atmosphere toward the zenith. Since
beam switching system is not equipped, we rapidly
scanned the antenna around the target CTA~102 in
azimuth and elevation direction, while recording the
total power. 
As a result, fluctuation of the
observed power due to the atmosphere can be
minimized and the pointing error can disappear. The
accuracy of the calibration was estimated to be 10\%.

\section{Results}

\begin{figure*}
  \centering
  \includegraphics[angle=0,width=14cm]{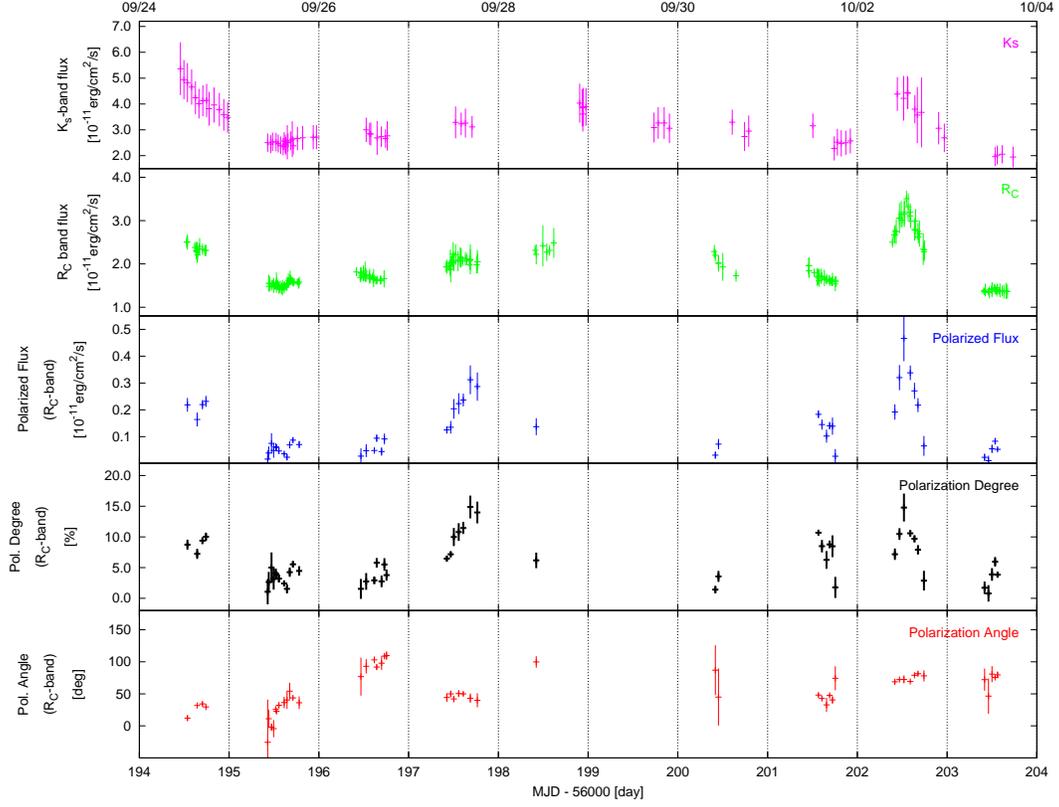}
  \caption{ Multi-wavelength light curves of CTA~102 from September 24 to Octobe
r 3 in 2012.
    Top panel: NIR ({\it K$_{\rm{s}}$}-band) flux.
    Second panel: optical ({\it R$_{\rm{C}}$}-band) flux.
    Third panel: polarized flux (PF) in the  {\it R$_{\rm{C}}$} band.
    Fourth panel: polarization degree (PD) in the  {\it R$_{\rm{C}}$} band.
    Bottom panel: polarization angle in the  {\it R$_{\rm{C}}$} band.
  }
  \label{fig:LC_MW}
\end{figure*}

Figure \ref{fig:LC_MW} shows temporal variations of the optical 
$R_{\rm C}$-band and NIR {\it K$_{\rm{s}}$}-band total fluxes, as well as  
those of the $R_{\rm C}$-band PD and PA during our dense monitoring of CTA~102, 
following intense $\gamma$-ray flare detected by {\it Fermi}-LAT. 
In the third panel of Figure 1, we also show a light curve of $R_{\rm C}$-band polarized flux (PF), which was calculated by
\begin{equation}
 PF = \frac{PD  \times F_{\rm R_{\rm C}}}{100},
\end{equation}
where $PD$ and $F_{\rm R_{\rm C}}$ are the measured polarization degree in unit of \% and total flux in the $R_{\rm C}$ band.
Except for some nights with bad weather, we were able to monitor CTA~102 continuously. We note that, although not shown here, light curves in other bands exhibited quite similar temporal variations to those in the $R_{\rm C}$ and $K_{\rm s}$ bands.

\begin{figure}
  \begin{center}
  \includegraphics[angle=0,width=8cm]{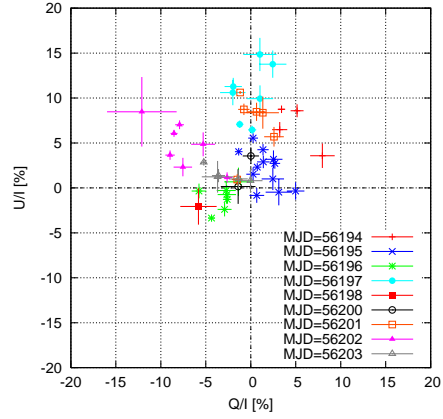}
  \caption{Intra-night and daily variations of the Stokes parameters 
    {\it Q} and {\it U} obtained by Kanata/HOWPol in our observation.
    Data points of the same color suggest the measurement on the same day.
    Dashed line (black) indicate the origin of the Stokes parameters.}
  \label{fig:QU}
  \end{center}
\end{figure}

First, the light curves in the $R_{\rm C}$ and
$K_{\rm s}$ bands showed a clear decay on MJD
56194. Given that {\it Fermi}-LAT detected a strong
$\gamma$-ray flare on MJD 56189, the observed
declining profile likely corresponds to the decay
phase of the bright $\gamma$-ray flare. After that,
both of the light curves showed a mild and
symmetrical enhancement peaking around MJD 56198
until the end of this follow-up observation, except
an intra-night strong flare on MJD 56202. In
addition, we can clearly see some remarkable
features in the daily polarimetric data. Namely,
possible PA swing from $\sim 0^{\circ}$ to $\sim
100^{\circ}$ was observed on MJD 56195 and 56196. To
check the temporal evolution of the PA, we plotted
the Stokes parameters $Q$ and $U$ observed during
these nights in Figure \ref{fig:QU}, together with
those of other nights. Again, gradual rotation of PA
was obviously seen, confirming the PA swing by $\sim
100^{\circ}$ during these two nights.

The continuous PA rotation was terminated by a sudden jump to
$\sim50^{\circ}$ on MJD 56197.  More strikingly, the PF rapidly
increased by three times from $\sim 0.1 \times 10^{-11}$ erg cm$^{-2}$
s$^{-1}$ to $\sim 0.3 \times 10^{-11}$ erg cm$^{-2}$ s$^{-1}$ without
an apparent increase of the total flux during the night.  This is the
first clear detection of a short-term (hours-scale) PF flare without
corresponding total flux enhancement.  Another interesting feature is
a strong intra-night flare detected on MJD 56202.  The isolated flare
showed smooth and symmetrical profile peaking around 13 UT, and the
total and polarized fluxes increased by factors of $\sim$1.5 and
$\sim$2.0, respectively. Notably, the PD evolved in a quite similar
way as the total flux, while the PA remained almost steady during the
flare (see also Figure \ref{fig:LC_MW} and Figure \ref{fig:QU}). The
PA measured on MJD 56202 is orthogonal to the jet direction observed
from VLBA observation \citep{2012arXiv1211.3606F}.  This is in
contrast with the case of AO 0235+164, in which the optical PA aligned
nearly the same direction as the jet during the flare in 2006 December
\citep{2008ApJ...672...40H}.

\begin{figure}
  \centering
  \includegraphics[angle=0,width=8cm]{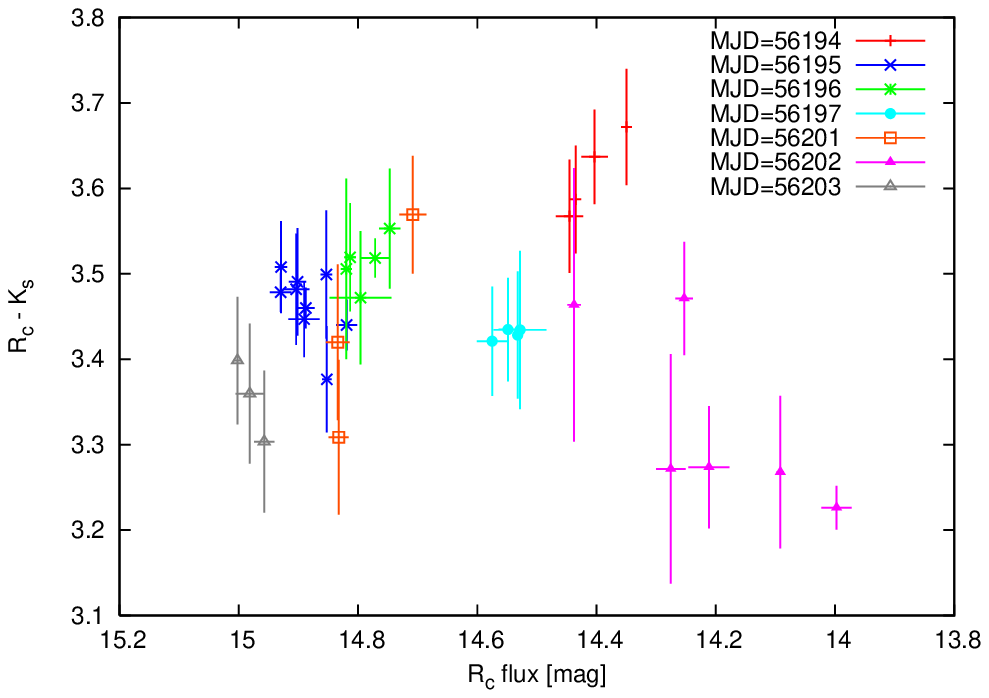}
  \caption{Intra-night and daily variations of {\it R$_{\rm{C}}-$K$_{\rm{s}}$} 
    color against {\it R$_{\rm{C}}$}-band flux. 
    Each mark show the same date as those used in Figure \ref{fig:QU}.
    }
  \label{fig:RmK}
\end{figure}

To see spectral evolution in the optical and NIR
bands, we plotted $R_{\rm C}-K_{\rm{s}}$ color
against $R_{\rm C}$-band magnitude in Figure \ref{fig:RmK}. 
As reported in \cite{2009AJ....138.1902O}, 
a ``redder when brighter'' trend is excepted in 
micro variability.
As a result, however, there is no clear correlation between
$R_{\rm C}-K_{\rm{s}}$ color and $R_{\rm C}$-band magnitude for a whole data.
Looking at the colors measured during our
observation other than MJD 56197 and 56202, the
source showed a hint of ``redder when brighter'' trend, as
is reported previously by \cite{2009AJ....138.1902O}. 

On the other hand, radio fluxes were almost constant
within 10\% during our monitoring and were
comparable to that measured in the quiescent state
\citep{1998A&AS..131..303S}. This is because the
emission region would be optically thick at 8.4
GHz. Actually, the constant radio flux during
optical and $\gamma$-ray flare was previously
observed for CTA~102 in 2006
\citep{2011A&A...531A..95F}.

\section{Discussion}
On MJD 56197, a transient PF flare without significant increase of the total flux was observed. This ``orphan PF flare" is interesting and explained in such way that a new emission component which is less luminous but possesses extremely high PD with respect to the long-term baseline component might have emerged suddenly. Large discrepancy of PA between MJD 56196 and 56197 ($\sim 100^{\circ}$ to $\sim 45^{\circ}$) supports the emergence of the new component. A careful inspection of the light curve showed that the source brightened only by $\sim$10\% in the total flux (which corresponds to $\sim 0.2 \times 10^{-11}$ erg cm$^{-2}$ s$^{-1}$) during the intra-night increase of PF from $0.1 \times 10^{-11}$ erg cm$^{-2}$ s$^{-1}$ to $0.3 \times 10^{-11}$ erg cm$^{-2}$ s$^{-1}$, implying that the
new component has the PD of $\sim$100\%. To overcome this unrealistic situation, we have to consider that
decrease of total flux of the long-term component occurred simultaneously with the emergence of the
highly-polarized new emission component and weakened increase of the total flux. In this case, the PD of
the new component depends on how much total flux of the long-term component decreases. For example, we
assume that gradual decrease of total flux of the long-term component by $0.2 \times 10^{-11}$ erg
cm$^{-2}$ s$^{-1}$ occurred simultaneously with gradual increase of the new component by $0.4 \times
10^{-11}$ erg cm$^{-2}$ s$^{-1}$ Then, the PD of the new component can be estimated as 50\%, which seems
reasonable compared to the above $\sim$100\% PD scenario, though still very high. We note that some
blazars such as OJ 287 and BL Lac exhibited similar orphan PF flares in longer day-month timescale
\citep{2011PASJ...63..639I}. It is currently unclear whether the same mechanism is responsible for such a
similar phenomenon with longer timescale. Anyway, our observation presented here is the first
measurement of the orphan PF flare in much shorter (intra-night) timescale, and further theoretical study is needed.

Smooth and continuous change of PA from $\sim0^{\circ}$ to $\sim100^{\circ}$ observed on MJD 56195 and 56196 suggests either the presence of helical magnetic field inside a jet \citep{2008Natur.452..966M} or global bending of a jet \citep{2010Natur.463..919A}. Note that the PD remained constant at $\sim 3$\% during the PA swing, which was also observed in both of the two scenarios from BL Lac (helical magnetic field) and 3C 279 (bent jet). The observed large PA swing rate of $\sim50^{\circ}$ per day is similar to the case of BL Lac and four times larger than that of 3C 279, implying that the helical magnetic field scenario might be preferred. On the other hand, if PA swing in the opposite direction is observed, the bent jet scenario would be validated.

Current Very Long Baseline Interferometry (VLBI) polarization observation is now feasible to examine magnetic field structure of the parsec-scale jet. Especially, measurement of Faraday rotation is recognized as a powerful tool to manifest the presence of helical magnetic field. When polarized radiation propagates within a magnetized plasma, the plane of linear polarization rotates and the rotation measure (RM, rotation of the polarization angle) depends on the line-of-sight component of the magnetic field \cite[see e.g.][]{Chen_1974}. Hence, if helical magnetic field is surrounding the jet, RM gradient across the jet should be observed \citep[e.g.][]{1993ASSL..103...15B}. Indeed, several authors \citep[e.g.,][]{2004MNRAS.351L..89G} claimed detection of RM gradient transverse to the VLBI jet in various blazars since the first discovery from 3C 273 by \cite{2002PASJ...54L..39A}, although the significance of these detections is still controversial \citep{2010ApJ...722L.183T}. Importantly, recent Very Long Baseline Array (VLBA) observation detected significant RM gradient from CTA 102 \citep{2012AJ....144..105H}. The presence of helical magnetic field is consistent with our observation of relatively fast PA swing over the two nights on MJD 56195 and 56196. Thus, optical polarimetry provides another powerful tool to claim the presence of helical magnetic field.

A prominent brightening was observed on MJD 56202, as is evident from the light curves of both the total flux and PF, with almost constant PA of $\sim 70^{\circ}$ (see Figure 1). This could be again explained by emergence of a new bright component with high PD with respect to the long-term baseline emission component. Comparison of the total and polarized fluxes between the beginning and peak of the flare allows us to estimate the PD of the new component. Namely, ratio of the PF increase by $\sim 0.35 \times 10^{-11}$ erg cm$^{-2}$ s$^{-1}$ with respect to the total-flux increase by $\sim 1.0 \times 10^{-11}$ erg cm$^{-2}$ s$^{-1}$ leads to the PD of $\sim$30--40\% for the new emission component. Such a high PD indicates presence of highly ordered magnetic field at the emission site, which might be generated through compression of turbulent magnetic field by shocks inside a jet as advocated by previous papers \citep[e.g., ][]{2008ApJ...672...40H}.

The optical PAs measured during the two flares on MJD 56197 and 56202 were $\sim 45^{\circ}$ and $\sim 70^{\circ}$, respectively. Compared to the VLBA jet image \citep[see e.g. Fig.A.15 of][for the jet image]{2013A&A...551A..32F}, we found that they are nearly orthogonal to the jet direction. Since magnetic field direction is in principle assumed to be perpendicular to PA, the measured PAs result in a claim of magnetic field orientation aligned with the jet. How is the highly-ordered magnetic field aligned with the jet generated? At first glance, the shock-in-jet scenario considered above seems unreasonable, because the magnetic field compressed by shocks propagating down the jet is aligned with a direction transverse to the jet. It should be noted that here magnetic field direction is inferred based on the assumption that it is orthogonal to the observed PA but recent theoretical work suggests that such an assumption is not correct for relativistically moving sources like AGN jets \citep{2005MNRAS.360..869L}. In particular,  \cite{2005MNRAS.360..869L} pointed out that toroidally dominated magnetic field is observed as poloidal in the observer's (rest) frame when ultra-relativistic jet with $\Gamma >>1$ is viewed at small angle to the line of sight ($\Gamma$ is bulk Lorentz factor of the jet). Given this complicated situation, the measured EVPAs significantly different from the jet direction can still be accounted for by the ``shock-in-jet" scenario.

It is interesting to compare the PAs observed during the two flares on MJD 56197 and 56202 with past observation, although only one observation by \cite{2008ApJ...672...40H} is found in the literature. The authors reported hours-scale short timescale polarimetric variability from BL Lac object AO 0235+164 during an outburst in 2006 December and found that the PAs tend to align with the jet direction around the maximum PD \citep[see Fig. 4 of][]{2008ApJ...672...40H}. This is quite different from our observation of perpendicular EVPAs to the jet direction. If relativistic effect hypothesis is true, uniform distribution of EVPA with respect to the jet direction would be found for hour-scale flares. In any case, the current samples are only two and further optical polarimetric observation is definitely needed. 

To summarize, we performed dense optical/IR photometric and polarimetric monitoring of CTA 102 following strong $\gamma$-ray flare in 2012 September with {\it OISTER} program. We found (i) smooth and gradual PA swing by $\sim 100$ deg over the two nights on MJD 56195 and 56196, (ii) orphan polarized flux flare on MJD 56197, (iii) significant brightening on MJD 56202 and (iv) grossly perpendicular PAs to the jet direction during the two flares. Combined with recent VLBA detection of RM gradient across the jet, we infer that helical magnetic field would be present at the emission region responsible for the long-term baseline component. The observed two flares can be explained by emergence of new emission component which possess highly-ordered magnetic field. Such a magnetic field configuration would be generated through compression by shocks propagating down the jet. The observed EVPAs perpendicular to the jet direction is not unreasonable given the effect of relativistically-moving radiation source.

\section{Acknowledgments}
This work is supported by Japan Society for the Promotion of Science (JSPS).
This work is also supported by Optical \& Near-infrared Astronomy Inter-University
Cooperation Program and Grants-in-Aid for Scientific Research 
(23340048, 24000004, 24244014, and 24840031) by the
Ministry of Education, Culture, Sports, Science and Technology of Japan.

\bibliographystyle{apj}

\begin{thebibliography}{36}
\expandafter\ifx\csname natexlab\endcsname\relax\def\natexlab#1{#1}\fi

\bibitem[{{Abdo} {et~al.}(2010{\natexlab{a}}){Abdo}, {Ackermann}, {Ajello},
  {Axelsson}, {Baldini}, {Ballet}, {Barbiellini}, {Bastieri}, {Baughman},
  {Bechtol}, \& et~al.}]{2010Natur.463..919A}
{Abdo}, A.~A., {et~al.} 2010{\natexlab{a}}, \nat, 463, 919

\bibitem[{{Abdo} {et~al.}(2010{\natexlab{b}}){Abdo}, {Ackermann}, {Ajello},
  {Antolini}, {Baldini}, {Ballet}, {Barbiellini}, {Bastieri}, {Bechtol},
  {Bellazzini}, {Berenji}, {Blandford}, {Bloom}, {Bonamente}, {Borgland},
  {Bouvier}, {Bregeon}, {Brez}, {Brigida}, {Bruel}, {Buehler}, {Burnett},
  {Buson}, {Caliandro}, {Cameron}, {Caraveo}, {Carrigan}, {Casandjian},
  {Cavazzuti}, {Cecchi}, {{\c C}elik}, {Chekhtman}, {Cheung}, {Chiang},
  {Ciprini}, {Claus}, {Cohen-Tanugi}, {Cominsky}, {Conrad}, {Costamante},
  {Cutini}, {Dermer}, {de Angelis}, {de Palma}, {Silva}, {Drell}, {Dubois},
  {Dumora}, {Farnier}, {Favuzzi}, {Fegan}, {Focke}, {Fortin}, {Frailis},
  {Fukazawa}, {Funk}, {Fusco}, {Gargano}, {Gasparrini}, {Gehrels}, {Germani},
  {Giebels}, {Giglietto}, {Giommi}, {Giordano}, {Glanzman}, {Godfrey},
  {Grenier}, {Grondin}, {Grove}, {Guiriec}, {Hadasch}, {Hayashida}, {Hays},
  {Healey}, {Horan}, {Hughes}, {Itoh}, {J{\'o}hannesson}, {Johnson}, {Johnson},
  {Kamae}, {Katagiri}, {Kataoka}, {Kawai}, {Kn{\"o}dlseder}, {Kuss}, {Lande},
  {Larsson}, {Latronico}, {Lemoine-Goumard}, {Longo}, {Loparco}, {Lott},
  {Lovellette}, {Lubrano}, {Madejski}, {Makeev}, {Massaro}, {Mazziotta},
  {McEnery}, {Michelson}, {Mitthumsiri}, {Mizuno}, {Moiseev}, {Monte},
  {Monzani}, {Morselli}, {Moskalenko}, {Mueller}, {Murgia}, {Nolan}, {Norris},
  {Nuss}, {Ohno}, {Ohsugi}, {Omodei}, {Orlando}, {Ormes}, {Ozaki}, {Panetta},
  {Parent}, {Pelassa}, {Pepe}, {Pesce-Rollins}, {Piron}, {Porter}, {Rain{\`o}},
  {Rando}, {Razzano}, {Reimer}, {Reimer}, {Ritz}, {Rodriguez}, {Romani},
  {Roth}, {Ryde}, {Sadrozinski}, {Sander}, {Scargle}, {Sgr{\`o}}, {Shaw},
  {Smith}, {Spandre}, {Spinelli}, {Starck}, {Strickman}, {Suson}, {Takahashi},
  {Takahashi}, {Tanaka}, {Thayer}, {Thayer}, {Thompson}, {Tibaldo}, {Torres},
  {Tosti}, {Tramacere}, {Uchiyama}, {Usher}, {Vasileiou}, {Vilchez}, {Vitale},
  {Waite}, {Wallace}, {Wang}, {Winer}, {Wood}, {Yang}, {Ylinen}, \&
  {Ziegler}}]{2010ApJ...722..520A}
---. 2010{\natexlab{b}}, \apj, 722, 520

\bibitem[{{Andruchow} {et~al.}(2003){Andruchow}, {Cellone}, {Romero},
  {Dominici}, \& {Abraham}}]{2003A&A...409..857A}
{Andruchow}, I., {Cellone}, S.~A., {Romero}, G.~E., {Dominici}, T.~P., \&
  {Abraham}, Z. 2003, \aap, 409, 857

\bibitem[{{Andruchow} {et~al.}(2011){Andruchow}, {Combi},
  {Mu{\~n}oz-Arjonilla}, {Romero}, {Cellone}, \&
  {Mart{\'{\i}}}}]{2011A&A...531A..38A}
{Andruchow}, I., {Combi}, J.~A., {Mu{\~n}oz-Arjonilla}, A.~J., {Romero}, G.~E.,
  {Cellone}, S.~A., \& {Mart{\'{\i}}}, J. 2011, \aap, 531, A38

\bibitem[{{Asada} {et~al.}(2002){Asada}, {Inoue}, {Uchida}, {Kameno},
  {Fujisawa}, {Iguchi}, \& {Mutoh}}]{2002PASJ...54L..39A}
{Asada}, K., {Inoue}, M., {Uchida}, Y., {Kameno}, S., {Fujisawa}, K., {Iguchi},
  S., \& {Mutoh}, M. 2002, \pasj, 54, L39

\bibitem[{{Blandford}(1993)}]{1993ASSL..103...15B}
{Blandford}, R. 1993, in Astrophysics and Space Science Library, Vol. 103,
  Astrophysics and Space Science Library, 15--33

\bibitem[{{Cellone} {et~al.}(2007){Cellone}, {Romero}, {Combi}, \&
  {Mart{\'{\i}}}}]{2007MNRAS.381L..60C}
{Cellone}, S.~A., {Romero}, G.~E., {Combi}, J.~A., \& {Mart{\'{\i}}}, J. 2007,
  \mnras, 381, L60

\bibitem[{{Chen}(1974)}]{Chen_1974}
{Chen}, F.~F. 1974, {Introduction to plasma physics} (New York: Plenum Press)

\bibitem[{{Fromm} {et~al.}(2011){Fromm}, {Perucho}, {Ros}, {Savolainen},
  {Lobanov}, {Zensus}, {Aller}, {Aller}, {Gurwell}, \&
  {L{\"a}hteenm{\"a}ki}}]{2011A&A...531A..95F}
{Fromm}, C.~M., {et~al.} 2011, \aap, 531, A95

\bibitem[{{Fromm} {et~al.}(2012){Fromm}, {Ros}, {Perucho}, {Savolainen},
  {Mimica}, {Kadler}, {Lobanov}, {Lister}, \& {Kovalev
  $\backslash$}}]{2012arXiv1211.3606F}
---. 2012, ArXiv e-prints

\bibitem[{{Fromm} {et~al.}(2013){Fromm}, {Ros}, {Perucho}, {Savolainen},
  {Mimica}, {Kadler}, {Lobanov}, {Lister}, {Kovalev}, \&
  {Zensus}}]{2013A&A...551A..32F}
---. 2013, \aap, 551, A32

\bibitem[{{Gabuzda} {et~al.}(2004){Gabuzda}, {Murray}, \&
  {Cronin}}]{2004MNRAS.351L..89G}
{Gabuzda}, D.~C., {Murray}, {\'E}., \& {Cronin}, P. 2004, \mnras, 351, L89

\bibitem[{{Hagen-Thorn} {et~al.}(2008){Hagen-Thorn}, {Larionov}, {Jorstad},
  {Arkharov}, {Hagen-Thorn}, {Efimova}, {Larionova}, \&
  {Marscher}}]{2008ApJ...672...40H}
{Hagen-Thorn}, V.~A., {Larionov}, V.~M., {Jorstad}, S.~G., {Arkharov}, A.~A.,
  {Hagen-Thorn}, E.~I., {Efimova}, N.~V., {Larionova}, L.~V., \& {Marscher},
  A.~P. 2008, \apj, 672, 40

\bibitem[{{Harris} \& {Roberts}(1960)}]{1960PASP...72..237H}
{Harris}, D.~E., \& {Roberts}, J.~A. 1960, \pasp, 72, 237

\bibitem[{{Hewitt} \& {Burbidge}(1989)}]{1989ApJS...69....1H}
{Hewitt}, A., \& {Burbidge}, G. 1989, \apjs, 69, 1

\bibitem[{{Hovatta} {et~al.}(2012){Hovatta}, {Lister}, {Aller}, {Aller},
  {Homan}, {Kovalev}, {Pushkarev}, \& {Savolainen}}]{2012AJ....144..105H}
{Hovatta}, T., {Lister}, M.~L., {Aller}, M.~F., {Aller}, H.~D., {Homan}, D.~C.,
  {Kovalev}, Y.~Y., {Pushkarev}, A.~B., \& {Savolainen}, T. 2012, \aj, 144, 105

\bibitem[{{Ikejiri} {et~al.}(2011){Ikejiri}, {Uemura}, {Sasada}, {Ito},
  {Yamanaka}, {Sakimoto}, {Arai}, {Fukazawa}, {Ohsugi}, {Kawabata}, {Yoshida},
  {Sato}, \& {Kino}}]{2011PASJ...63..639I}
{Ikejiri}, Y., {et~al.} 2011, \pasj, 63, 639

\bibitem[{{Kawabata} {et~al.}(2008){Kawabata}, {Nagae}, {Chiyonobu}, {Tanaka},
  {Nakaya}, {Suzuki}, {Kamata}, {Miyazaki}, {Hiragi}, {Miyamoto}, {Yamanaka},
  {Arai}, {Yamashita}, {Uemura}, {Ohsugi}, {Isogai}, {Ishitobi}, \&
  {Sato}}]{2008SPIE.7014E.151K}
{Kawabata}, K.~S., {et~al.} 2008, in Society of Photo-Optical Instrumentation
  Engineers (SPIE) Conference Series, Vol. 7014, Society of Photo-Optical
  Instrumentation Engineers (SPIE) Conference Series

\bibitem[{{Kotani} {et~al.}(2005){Kotani}, {Kawai}, {Yanagisawa}, {Watanabe},
  {Arimoto}, {Fukushima}, {Hattori}, {Inata}, {Izumiura}, {Kataoka}, {Koyano},
  {Kubota}, {Kuroda}, {Mori}, {Nagayama}, {Ohta}, {Okada}, {Okita}, {Sato},
  {Serino}, {Shimizu}, {Shimokawabe}, {Suzuki}, {Toda}, {Ushiyama}, {Yatsu},
  {Yoshida}, \& {Yoshida}}]{2005NCimC..28..755K}
{Kotani}, T., {et~al.} 2005, Nuovo Cimento C Geophysics Space Physics C, 28,
  755

\bibitem[{{Landolt}(1992)}]{1992AJ....104..340L}
{Landolt}, A.~U. 1992, \aj, 104, 340

\bibitem[{{Larionov} {et~al.}(2012){Larionov}, {Blinov}, \&
  {Jorstad}}]{2012ATel.4397....1L}
{Larionov}, V., {Blinov}, D., \& {Jorstad}, S. 2012, The Astronomer's Telegram,
  4397, 1

\bibitem[{{Lyutikov} {et~al.}(2005){Lyutikov}, {Pariev}, \&
  {Gabuzda}}]{2005MNRAS.360..869L}
{Lyutikov}, M., {Pariev}, V.~I., \& {Gabuzda}, D.~C. 2005, \mnras, 360, 869

\bibitem[{{Marscher} {et~al.}(2008){Marscher}, {Jorstad}, {D'Arcangelo},
  {Smith}, {Williams}, {Larionov}, {Oh}, {Olmstead}, {Aller}, {Aller},
  {McHardy}, {L{\"a}hteenm{\"a}ki}, {Tornikoski}, {Valtaoja}, {Hagen-Thorn},
  {Kopatskaya}, {Gear}, {Tosti}, {Kurtanidze}, {Nikolashvili}, {Sigua},
  {Miller}, \& {Ryle}}]{2008Natur.452..966M}
{Marscher}, A.~P., {et~al.} 2008, \nat, 452, 966

\bibitem[{{Miller} \& {Noble}(1996)}]{1996ASPC..110...17M}
{Miller}, H.~R., \& {Noble}, J.~C. 1996, in Astronomical Society of the Pacific
  Conference Series, Vol. 110, Blazar Continuum Variability, ed. H.~R.
  {Miller}, J.~R. {Webb}, \& J.~C. {Noble}, 17

\bibitem[{{Orienti} \& {D'Ammando}(2012)}]{2012ATel.4409....1O}
{Orienti}, M., \& {D'Ammando}, F. 2012, The Astronomer's Telegram, 4409, 1

\bibitem[{{Osterman Meyer} {et~al.}(2009){Osterman Meyer}, {Miller},
  {Marshall}, {Ryle}, {Aller}, {Aller}, \& {Balonek}}]{2009AJ....138.1902O}
{Osterman Meyer}, A., {Miller}, H.~R., {Marshall}, K., {Ryle}, W.~T., {Aller},
  H., {Aller}, M., \& {Balonek}, T. 2009, \aj, 138, 1902

\bibitem[{{Racine}(1970)}]{1970ApJ...159L..99R}
{Racine}, R. 1970, \apjl, 159, L99

\bibitem[{{Sasada} {et~al.}(2008){Sasada}, {Uemura}, {Arai}, {Fukazawa},
  {Kawabata}, {Ohsugi}, {Yamashita}, {Isogai}, {Sato}, \&
  {Kino}}]{2008PASJ...60L..37S}
{Sasada}, M., {et~al.} 2008, \pasj, 60, L37

\bibitem[{{Schlafly} \& {Finkbeiner}(2011)}]{2011ApJ...737..103S}
{Schlafly}, E.~F., \& {Finkbeiner}, D.~P. 2011, \apj, 737, 103

\bibitem[{{Schulz} \& {Lenzen}(1983)}]{1983A&A...121..158S}
{Schulz}, A., \& {Lenzen}, R. 1983, \aap, 121, 158

\bibitem[{{Skrutskie} {et~al.}(2006){Skrutskie}, {Cutri}, {Stiening},
  {Weinberg}, {Schneider}, {Carpenter}, {Beichman}, {Capps}, {Chester},
  {Elias}, {Huchra}, {Liebert}, {Lonsdale}, {Monet}, {Price}, {Seitzer},
  {Jarrett}, {Kirkpatrick}, {Gizis}, {Howard}, {Evans}, {Fowler}, {Fullmer},
  {Hurt}, {Light}, {Kopan}, {Marsh}, {McCallon}, {Tam}, {Van Dyk}, \&
  {Wheelock}}]{2006AJ....131.1163S}
{Skrutskie}, M.~F., {et~al.} 2006, \aj, 131, 1163

\bibitem[{{Stanghellini} {et~al.}(1998){Stanghellini}, {O'Dea}, {Dallacasa},
  {Baum}, {Fanti}, \& {Fanti}}]{1998A&AS..131..303S}
{Stanghellini}, C., {O'Dea}, C.~P., {Dallacasa}, D., {Baum}, S.~A., {Fanti},
  R., \& {Fanti}, C. 1998, \aaps, 131, 303

\bibitem[{{Taylor} \& {Zavala}(2010)}]{2010ApJ...722L.183T}
{Taylor}, G.~B., \& {Zavala}, R. 2010, \apjl, 722, L183

\bibitem[{{Watanabe} {et~al.}(2012){Watanabe}, {Takahashi}, {Sato}, {Watanabe},
  {Fukuhara}, {Hamamoto}, \& {Ozaki}}]{2012SPIE.8446E..2OW}
{Watanabe}, M., {Takahashi}, Y., {Sato}, M., {Watanabe}, S., {Fukuhara}, T.,
  {Hamamoto}, K., \& {Ozaki}, A. 2012, in Society of Photo-Optical
  Instrumentation Engineers (SPIE) Conference Series, Vol. 8446, Society of
  Photo-Optical Instrumentation Engineers (SPIE) Conference Series

\bibitem[{{Yanagisawa} {et~al.}(2006){Yanagisawa}, {Shimizu}, {Okita},
  {Nagayama}, {Sato}, {Koyano}, {Okada}, {Iwata}, {Uraguchi}, {Watanabe},
  {Yoshida}, {Okumura}, {Nakaya}, \& {Yamamuro}}]{2006SPIE.6269E.118Y}
{Yanagisawa}, K., {et~al.} 2006, in Society of Photo-Optical Instrumentation
  Engineers (SPIE) Conference Series, Vol. 6269, Society of Photo-Optical
  Instrumentation Engineers (SPIE) Conference Series

\bibitem[{{Yanagisawa} {et~al.}(2008){Yanagisawa}, {Okita}, {Shimizu},
  {Otsuka}, {Nagayama}, {Iwata}, {Ozaki}, {Yoshida}, {Nakaya}, {Tajitsu},
  {Okumura}, \& {Yamamuro}}]{2008SPIE.7014E.106Y}
{Yanagisawa}, K., {et~al.} 2008, in Society of Photo-Optical Instrumentation
  Engineers (SPIE) Conference Series, Vol. 7014, Society of Photo-Optical
  Instrumentation Engineers (SPIE) Conference Series

\end{thebibliography}

\end{document}